\documentclass[aps,prd,twocolumn,groupedaddress,showpacs,showkeys,nofootinbib,amssymb]{revtex4}
\usepackage[latin1]{inputenc}
\usepackage{graphicx}
\usepackage[english]{babel}
\usepackage{amsmath}
\usepackage{amssymb}
\usepackage{amsfonts}

\begin{document}
\def\pp{{\, \mid \hskip -1.5mm =}}
\def\cL{{\cal L}}
\def\beq{\begin{equation}}
\def\eneq{\end{equation}}
\def\bea{\begin{eqnarray}}
\def\enea{\end{eqnarray}}
\def\tr{{\rm tr}\, }
\def\nn{\nonumber \\}
\def\e{{\rm e}}

\title{\textbf{Correspondence between Jordan-Einstein frames and Palatini-metric formalisms}}

\author{Salvatore Capozziello$^{\ast,\S}$, Farhad Darabi$^\bullet$, Daniele Vernieri$^\ast$}

\affiliation{\it $^\ast$Dipartimento di Scienze Fisiche,
Università di Napoli {}``Federico II'', $^\S$INFN Sez. di Napoli,
Compl. Univ. di
Monte S. Angelo, Edificio G, Via Cinthia, I-80126, Napoli, Italy \\
$^\bullet$Department of Physics, Azarbaijan University of Tarbiat Moallem, Tabriz 53741-161, Iran\\
Research Institute for Astronomy and Astrophysics of Maragha (RIAAM), Maragha 55134-441, Iran}

\date{\today}

\begin{abstract}
We discuss the conformal  symmetry between Jordan and Einstein
frames considering their relations with  the metric and Palatini
formalisms for modified gravity. Appropriate conformal
transformations are taken into account leading to the evident
connection between the gravitational actions in the two mentioned
frames and the Hilbert-Einstein action with a cosmological
constant. We show that the apparent differences between Palatini
and metric formalisms strictly depend on the representation while
the number of degrees of freedom is preserved. This means that the
dynamical content of both formalism is identical.
\end{abstract}


 \maketitle

\section{introduction}
\label{1}

The recent interest in investigating alternative theories of
gravity has  arisen from cosmology, quantum field theory and
Mach's principle. The initial singularity, flatness and horizon
problems \cite{guth} indicate that the standard cosmological
model, based on general relativity (GR) and the particle standard
model \cite{wald, wein, misn}, fails in describing the Universe at
extreme regimes. Besides, GR does not work as a fundamental theory
capable of giving a quantum description of spacetime
\cite{birrell}. For these reasons and due to the lack of a
definitive quantum gravity theory, alternative theories of
gravitation have been pursued in order to attempt an at least
semi-classical approach to quantization and to early universe
shortcomings.

In particular extended theories of gravity (ETGs)
\cite{mauro,farasot,libroSaFe} take into account the problem of
gravitational interaction correcting and enlarging the Einstein
theory  by the introduction of non-minimally coupled scalar fields
and higher-order terms in curvature invariants.  The idea to
extend GR is fruitful and economic with respect to several
attempts which try to solve problems by adding new and, sometime,
unjustified new ingredients in order to give a self-consistent
picture of the cosmic and quantum dynamics (e.g. dark energy and
dark matter up to now not detected at fundamental level). In
particular, such an approach  `naturally' reproduce inflationary
behaviors in early epochs  and is capable of  matching with
several astrophysical observations. Besides, the present-day
observed accelerated expansion of Hubble flow and the missing
matter of astrophysical large-scale structures could be explained
by changing the gravitational sector, i.e. the lhs of the field
equations \cite{JCAP}. The alternative philosophy is to add new
cosmic fluids (new components in the rhs of the field equations)
which should give rise to clustered structures (dark matter) or to
accelerated dynamics (dark energy) thanks to exotic equations of
state. In particular, relaxing the hypothesis that gravitational
Lagrangian has to be a linear function of the Ricci curvature
scalar $R$, as in the Hilbert-Einstein formulation, one can take
into account, as a minimal extension, an effective action where
the gravitational Lagrangian is a generic $f(R)$ function
\cite{f(R)-cosmo,f(R)-cosmo1,f(R)-cosmo2,f(R)-cosmo3,voll,voll1}.

Moreover one can consider  actions where  scalar field are
non-minimally coupled to  gravity \cite{cimall}  as generalization
of the Brans-Dicke theory  \cite{brans}. Through the conformal
transformations, it is possible to show that any higher-order or
scalar-tensor theory, in absence of ordinary matter, e.g. a
perfect fluid, is conformally equivalent to an Einstein theory
plus minimally coupled scalar fields. In principle, the converse
is also true:  we can transform  standard Einstein gravity plus
minimally coupled scalar fields into a non-minimally coupled
scalar-tensor theory.

Conformal transformations can be useful to point out common
features between Palatini and metric approaches to gravitational
interaction.  The fundamental idea of the Palatini formalism is to
consider the connection $\Gamma$, entering the definition of the
Ricci tensor, to be independent of the metric $g$ defined on the
spacetime manifold $\mathcal{M}$. Conceptually, this means that
geodesic and causal structures on  $\mathcal{M}$ can be
disentangled \cite{FoP}. The Palatini formulation for the standard
Hilbert-Einstein theory results to be equivalent to the purely
metric theory: this follows from the fact that the field equations
for the connection $\Gamma$, firstly considered to be independent
of the metric, give the Levi-Civita connection of the metric $g$.
As a consequence, there is no reason to impose the Palatini
variational principle in the standard Hilbert-Einstein theory
instead of the metric variational principle. The situation changes
if we consider  ETGs, depending on functions of curvature
invariants, as $f(R)$, or non-minimally coupled to some scalar
field. In these cases, the Palatini and the metric variational
principles provide different field equations and the theories thus
derived seem to differ \cite{magn, ferr}. This status of art is
not comfortable since  dynamics and its predictions should not
depend on the representation. In fact, it is well known that
several astrophysical and cosmological observations can be well
interpreted in a formalism and not in the other and viceversa
\cite{mauro,farasot}. This shortcoming can be partially removed by
investigating how Palatini and metric formalisms are related by
conformal transformations.

In this paper, we  discuss the correspondence between
Jordan-Einstein frames and Palatini-metric formalisms pointing out
how Lagrangians can be transformed between each other and that the
number of degrees of freedom is preserved.

In Sec.\ref{2} we discuss the conformal symmetry between Jordan
and Einstein frames. In Sec.\ref{3} we introduce metric and
Palatini formalisms for some ETGs, and in Sec.\ref{4}, we use some
appropriate transformations from Jordan to Einstein frames in view
to compare  Palatini and metric formalisms. Conclusions and some
physical considerations  are given in Sec.\ref{5}.

\section{Conformal symmetry between Jordan and Einstein frames} \label{2}

The general form of the action in four dimensions when there is a
nonstandard coupling between a scalar field and the geometry is
\begin{equation}\label{eq01}
{\cal S}=\int d^4x \sqrt{-g}\left(F(\phi)R+\frac{1}{2}g^{\mu \nu}\phi_{;\mu}\phi_{;\nu}-V(\phi)\right),
\end{equation}
where $R$ is the Ricci scalar, $V(\phi)$ and $F(\phi)$ are
functions describing the effective potential and the coupling of
$\phi$ with gravity, respectively\footnote{The metric signature is
(- + + +) and Planck units are adopted.}. This form of the action
or the related Lagrangian density is usually referred to the {\it
Jordan frame}. The variation with respect to the metric $g_{\mu
\nu}$ gives  the generalized Einstein equations
\begin{equation}\label{eq02}
F(\phi)G_{\mu \nu}=-\frac{1}{2}T_{\mu \nu}-g_{\mu \nu}\square_{\Gamma}F(\phi)+F(\phi)_{;\mu\nu},
\end{equation}
where $\square_{\Gamma}$ is the d'Alembert operator with respect to the connection
$\Gamma$, $G_{\mu \nu}$ is the standard Einstein tensor
\begin{equation}\label{eq03}
G_{\mu \nu}=R_{\mu \nu}-\frac{1}{2}R g_{\mu \nu},
\end{equation}
and $T_{\mu \nu}$ is the energy-momentum tensor of the scalar field
\begin{equation}\label{eq04}
T_{\mu \nu}=\phi_{;\mu}\phi_{;\nu}-\frac{1}{2}g_{\mu \nu}\phi_{;\alpha}\phi^{;\alpha}+g_{\mu \nu}V(\phi).
\end{equation}
 The variation with respect to $\phi$ leads to the Klein-Gordon
equation
\begin{equation}\label{eq05}
\square_{\Gamma}\phi-RF_{\phi}(\phi)+V_{\phi}(\phi)=0,
\end{equation}
where $F_{\phi}=\frac{dF(\phi)}{d\phi},
V_{\phi}(\phi)=\frac{dV(\phi)}{d\phi} $.
Let us  consider now a
conformal transformation on the metric $g_{\mu \nu}$
\begin{equation}\label{eq06}
\bar{g}_{\mu \nu}=e^{2\omega}g_{\mu \nu},
\end{equation}
with the conformal factor $e^{2\omega}$. The Lagrangian density in
(\ref{eq01}) becomes \bea    \label{eq07}
\sqrt{-g}\left(FR+\frac{1}{2}g^{\mu \nu}\phi_{;\mu}\phi_{;\nu}-V(\phi)\right)=\sqrt{-\bar{g}}e^{-2\omega}\left(F\bar{R}+ \right.\nonumber \\
\left.-6F\square_{\bar{\Gamma}}\omega-6F\omega_{;\alpha}\omega^{;\alpha}+\frac{1}{2}\bar{g}^{\mu
\nu}\phi_{;\mu}\phi_{;\nu}-e^{-2\omega}V\right), \enea where
$\bar{R}, \bar{\Gamma}$ and $\square_{\bar{\Gamma}}$ are the
corresponding quantities with respect to the metric $\bar{g}_{\mu
\nu}$ and connection $\bar{\Gamma}$, respectively.  If we require
that the new Lagrangian, in terms of $\bar{g}_{\mu \nu}$,  appears
as a standard Einstein theory, the conformal factor has to be
related to $F$ as
\begin{equation}\label{eq08}
e^{2\omega}=2F.
\end{equation}
Using this relation, the Lagrangian  (\ref{eq07}) becomes
\bea\label{eq09} \sqrt{-g}\left(FR+\frac{1}{2}g^{\mu
\nu}\phi_{;\mu}\phi_{;\nu}-V(\phi)\right)=
\sqrt{-\bar{g}}\left(\frac{1}{2}\bar{R}+3\square_{\bar{\Gamma}}\omega+\right. \nonumber \\
\left.+\frac{3F_{\phi}^2-F}{4F^2}\phi_{;\alpha}\phi^{;\alpha}-\frac{V}{4F^2}\right)
\enea By introducing a new scalar field $\bar{\phi}$ and the
related potential $\bar{V}$ defined as
\begin{equation}\label{eq10}
\bar{\phi}_{;\alpha}=\sqrt{\frac{3F_{\phi}^2-F}{4F^2}}\phi_{;\alpha},
\:\:\:\:\: \bar{V}(\bar{\phi}(\phi))=\frac{V}{4F^2},
\end{equation}
we obtain\footnote{Note that the divergence-type
term $3\square_{\bar{\Gamma}}\omega$ appearing in the Lagrangian density is not considered \cite{CRM}.}
\bea    \label{eq11}
\sqrt{-g}\left(FR+\frac{1}{2}g^{\mu \nu}\phi_{;\mu}\phi_{;\nu}-
V(\phi)\right)=\sqrt{-\bar{g}}\left(\frac{1}{2}\bar{R}\right.+ \nonumber \\
\left.+\frac{1}{2}\bar{\phi}_{;\alpha}\bar{\phi}^{;\alpha}-\bar{V}\right),
\enea where the r.h.s. is the usual Einstein-Hilbert Lagrangian
density subject to the metric $\bar{g}_{\mu \nu}$, plus the
standard Lagrangian density of the scalar field $\bar{\phi}$. This
form of the Lagrangian density is usually referred to the {\it
Einstein frame}.  Therefore, we realize that any non-minimally
coupled theory of gravity with scalar field, in  absence of
ordinary matter, is conformally equivalent to the standard
Einstein gravity coupled with scalar field provided we use the
conformal transformation (\ref{eq08}) together with the
definitions (\ref{eq10}). The converse is also true: for a given
$F(\phi)$, such that
\begin{equation}
{3F_{\phi}^2-F}>0\,,
\end{equation}
that means the Hessian determinant is non singular and the
coupling has the right signature,  we can transform a standard
Einstein theory into a nonstandard coupled theory. This has an
important meaning: if we are able to solve the field equations
within the framework of standard Einstein gravity coupled with a
scalar field subject to a given potential, we are be able, in
principle,  to get  solutions for the class of nonstandard coupled
theories, with the coupling $F(\phi)$, through the conformal
transformation defined by (\ref{eq08}), the only constraint being
the second equation of (\ref{eq10}). This statement is exactly
what we mean as the {\it conformal equivalence between Jordan and
Einstein frames}. However, this mathematical equivalence does not
imply directly the physical equivalence of the two frames.
Examples in this sense can be found in \cite{prado,cnot,CNO}.

\section{Metric and Palatini formalism for modified gravity}
\label{3}
The action in the metric formalism for $f(R)$ gravity
takes the form
\begin{equation}\label{eq12'}
{\cal S}=\int_m d^4x \sqrt{-g}f({R}).
\end{equation}
In the metric formalism, the variation of the action is
accomplished with respect to the metric. One can show that this
action dynamically corresponds to an action of non-minimally
coupled gravity with a new scalar field having no kinetic term. By
introducing a new auxiliary field $\chi$, the dynamically
equivalent action can be rewritten as \cite{farasot, Sotiriou}
\begin{equation}\label{eq13'}
{\cal S}=\int_m d^4x
\sqrt{-g}(f(\chi)+f^\prime(\chi)(R-\chi)).
\end{equation}
Variation with respect to $\chi$ yields the equation
\begin{equation}\label{eq14'}
f^{\prime \prime}(\chi)({R}-\chi)=0.
\end{equation}
Therefore, $\chi=R$, if $f''(\chi)\neq0$,  reproduces the action
(\ref{eq12'}). Redefining the field $\chi$ by $\phi =
f^\prime(\chi)$ and introducing  the potential \beq
V(\phi)=\chi(\phi)\phi-f(\chi(\phi)), \eneq the action
(\ref{eq13'}) takes the form
\begin{equation}\label{eq15'}
{\cal S}=\int_m d^4x \sqrt{-g}(\phi{R}-V(\phi)),
\end{equation}
that is the Jordan frame representation  of the action of a
Brans-Dicke theory with Brans-Dicke parameter $\omega_0=0$, known
as O'Hanlon action in metric formalism.

Beside the metric formalism  in which the variation of the action
is done with respect to the metric, the Einstein equations can be
derived as well using the Palatini formalism, i.e. the variation
with respect to the metric is independent of the variation with
respect to the connection. The Riemann tensor and the Ricci tensor
are also constructed with the independent connection and the
metric is not needed to obtain the latter from the former. So, in
order to make a difference with metric formalism, we shall use
${\cal R}_{\mu \nu}$ and ${\cal R}$ instead of $R_{\mu \nu}$ and
$R$, respectively. In the standard Einstein-Hilbert action there
is no specific difference between these two formalisms. However,
once we generalize the action to depend on a generalized form of
the Ricci scalar they are no longer the same.

We briefly review the $f({\cal R})$ gravity in Palatini formalism
and show how it corresponds to a Brans-Dicke theory
\cite{mauro,farasot}. The action in the Palatini formalism with no
matter is written as
\begin{equation}\label{eq12}
{\cal S}=\int_p d^4x \sqrt{-g}f({\cal R}).
\end{equation}
Varying the action (\ref{eq12}) independently with respect to the
metric and the connection, respectively, and using the formula
\begin{equation}\label{eq13}
\delta{\cal R}_{\mu \nu}=\bar{\nabla}_{\lambda}\delta
\Gamma^{\lambda}_{\mu \nu}-\bar{\nabla}_{\nu}\delta
\Gamma^{\lambda}_{\mu \lambda},
\end{equation}
yields
\begin{equation}\label{eq14}
f^{\prime}({\cal R}){\cal R}_{(\mu \nu)}-\frac{1}{2}f({\cal
R})g_{\mu \nu}=0,
\end{equation}
\begin{equation}\label{eq15}
\bar{\nabla}_{\lambda}(\sqrt{-g}f^{\prime}({\cal R})g^{\mu
\nu})-\bar{\nabla}_{\sigma}(\sqrt{-g}f^{\prime}({\cal R})g^{\sigma
(\mu })\delta_{\lambda}^{\nu)}=0,
\end{equation}
where $\bar{\nabla}$ denotes the covariant derivative defined with
the independent connection $\Gamma^{\lambda}_{\mu \nu}$ and $(\mu
\nu)$ denotes symmetrization over the indices $\mu, \nu$. Taking
the trace of Eq. (\ref{eq15}) gives
\begin{equation}\label{eq16}
\bar{\nabla}_{\sigma}(\sqrt{-g}f^{\prime}({\cal R})g^{\sigma \mu
})=0,
\end{equation}
by which the field equation (\ref{eq15}) becomes
\begin{equation}\label{eq17}
\bar{\nabla}_{\lambda}(\sqrt{-g}f^{\prime}({\cal R})g^{\mu
\nu})=0.
\end{equation}
One may obtain some useful manipulations of the field equations.
Taking the trace of Eq. (\ref{eq14}) yields an algebraic equation
for ${\cal R}$
\begin{equation}\label{eq18}
f^{\prime}({\cal R}){\cal R}-{2}f({\cal R})=0.
\end{equation}
One can define a metric conformal to $g_{\mu \nu}$ as
\begin{equation}\label{eq19}
h_{\mu \nu}=f^\prime({\cal R})g_{\mu \nu},
\end{equation}
for which it is easily obtained that
\begin{equation}\label{eq20}
\sqrt{-h}h^{\mu \nu}=\sqrt{-g}f^\prime({\cal R})g^{\mu
\nu}.
\end{equation}
Eq. (\ref{eq17}) is then the compatibility condition of the metric
$h_{\mu \nu}$ with the connection $\Gamma^{\lambda}_{\mu \nu}$ and
can be solved algebraically to give the Levi-Civita connection
\begin{equation}\label{eq21}
\Gamma^{\lambda}_{\mu \nu}=h^{\lambda \sigma}(\partial_{\mu} h_{\nu
\sigma}+\partial_{\nu} h_{\mu \sigma}-\partial_{\sigma} h_{\mu
\nu}).
\end{equation}
Under  conformal transformation (\ref{eq19}), the Ricci tensor and
its contracted form with $g^{\mu \nu}$ become, respectively,
\bea
\label{eq22} {\cal R}_{\mu \nu}=R_{\mu
\nu}&+&\frac{3}{2}\frac{1}{(f^\prime({\cal
R}))^2}(\nabla_{\mu}f^\prime({\cal R}))(\nabla_{\nu}f^\prime({\cal R}))+ \nonumber \\
&-&\frac{1}{(f^\prime({\cal
R}))}(\nabla_{\mu}\nabla_{\nu}-\frac{1}{2}g_{\mu
\nu}\Box)f^\prime({\cal R}),
\enea
\begin{equation}\label{eq23}
{\cal R}=R+\frac{3}{2}\frac{1}{(f^\prime({\cal
R}))^2}(\nabla_{\mu}f^\prime({\cal R}))(\nabla^{\mu}f^\prime({\cal
R}))+\frac{3}{(f^\prime({\cal R}))}\Box f^\prime({\cal
R}).
\end{equation}
Note the difference between ${\cal R}$ and the Ricci scalar of
$h_{\mu \nu}$ is due to the fact that $g^{\mu \nu}$ is used here
for the contraction of ${\cal R}_{\mu \nu}$. Now, by introducing a
new auxiliary field $\chi$, the dynamically equivalent action is
rewritten as \cite{farasot, Sotiriou}
\begin{equation}\label{eq24}
{\cal S}=\int_P d^4x
\sqrt{-g}(f(\chi)+f^\prime(\chi)({\cal R}-\chi)).
\end{equation}
Variation with respect to $\chi$ yields the equation
\begin{equation}\label{eq25}
f^{\prime \prime}(\chi)({\cal R}-\chi)=0.
\end{equation}
Redefining the field $\chi$ by $\phi = f^\prime(\chi)$ and
introducing \beq V(\phi)=\chi(\phi)\phi-f(\chi(\phi)), \eneq with
the same  request made in the metric formalism, $f''(\chi)\neq0$
which implies ${\cal R}=\chi$, the action (\ref{eq24}) takes the
form
\begin{equation}\label{eq26}
{\cal S}=\int_P d^4x \sqrt{-g}(\phi{\cal R}-V(\phi)).
\end{equation}
Now, we may use $\phi = f^\prime(\chi)$ in Eq. (\ref{eq23}) to write down
${\cal R}$ in terms of $R$ in the action (\ref{eq26}). This leads, modulo a surface term, to
\begin{equation}\label{eq27}
{\cal S}=\int_P d^4x \sqrt{-g}\left(\phi
R+\frac{3}{2\phi}\nabla_{\mu}\phi \nabla^{\mu}\phi-V(\phi)\right).
\end{equation}
This is the action in Palatini formalism which corresponds to a
Brans-Dicke theory  with  $\omega =-\frac{3}{2}$. These results
are well known. How aim is now to show that the dynamical
information in both metric and Palatini formalisms is the same and
that the number of degrees of freedom is preserved.

\section{Transformation from Jordan to Einstein frames}
\label{4} Let us now use some appropriate transformations to
manipulate  the actions (\ref{eq15'}) and (\ref{eq27}),
respectively in metric and Palatini formalisms, from the Jordan to
the Einstein frame. Comparison of the action (\ref{eq15'}) with
(\ref{eq27}) reveals, as we have already specified, that the
former is the action of a Brans-Dicke theory with $\omega_0=0$. We
first define the conformal metric $\bar{g}_{\mu \nu}=\Phi g_{\mu
\nu}$ and perform a conformal transformation along with $\Phi=R$
assuming the scalar field definition
$\Phi=\exp{(\frac{\sqrt{3}}{2}\varphi)}$. One  therefore obtains
an action describing Einstein gravity minimally coupled to a
scalar field, that is \cite{Faraoni,Faraoni1}
\begin{equation}\label{eq27'}
{\cal S}=\int_m d^4x \sqrt{-\bar{g}}\left(
\bar{R}-\frac{1}{2}\nabla_{\mu}\varphi \nabla^{\mu}\varphi-V(\varphi)\right),
\end{equation}
where $\bar{R}$ is the Ricci scalar of the metric $\bar{g}$. This action
is now said to be written in the Einstein frame.

On the other hand, if we redefine the scalar field $\phi$  as the
new field
\begin{equation}\label{eq28}
\sigma=2\sqrt{3\phi},
\end{equation}
the Brans-Dicke action (\ref{eq27}) then becomes
\begin{equation}\label{eq29}
{\cal S}=\int_P d^4x \sqrt{-g}\left(F(\sigma)
R+\frac{1}{2}g^{\mu \nu}\sigma_{;\mu}\sigma_{;\nu}-V(\sigma)\right),
\end{equation}
where
\begin{equation}\label{eq30}
F(\sigma)=\frac{1}{12}\sigma^2.
\end{equation}
This action is now exactly the same as (\ref{eq01}) in the Jordan
frame in which $\phi$ is replaced by $\sigma$. However, it is
worth noticing that action (\ref{eq29}) is derived from the
Palatini formalism while  (\ref{eq01}) is defined in the metric
formalism. Therefore, with a similar procedure for the field
$\sigma$ we can write  \bea\label{eq31}
\sqrt{-g}\left(F(\sigma)R+\frac{1}{2}g^{\mu
\nu}\sigma_{;\mu}\sigma_{;\nu}-V(\sigma)\right)=
\sqrt{-\bar{g}}\left(\frac{1}{2}\bar{R}+ \right.\nonumber \\
+\left.\frac{1}{2}\bar{\sigma}_{;\alpha}\bar{\sigma}^{;\alpha}
-\bar{V}\right) \enea
where
\begin{equation}\label{eq32}
\bar{\sigma}_{;\alpha}=\sqrt{\frac{3F_{\sigma}^2-F}{4F^2}}\sigma_{;\alpha},
\:\:\:\:\: \bar{V}(\bar{\sigma}(\sigma))=\frac{V}{4F^2}.
\end{equation}
and \beq F_\sigma=\frac{dF(\sigma)}{d\sigma}. \eneq Substituting
$F(\sigma)$ in the definition of $\bar{\sigma}_{;\alpha}$  leads
to zero kinetic term for this field and finally we obtain
\begin{equation}\label{eq33}
\sqrt{-g}\left(F(\sigma)R+\frac{1}{2}g^{\mu \nu}\sigma_{;\mu}\sigma_{;\nu}-V(\sigma)\right)=
\sqrt{-\bar{g}}\left(\frac{1}{2}\bar{R}
-\bar{V}\right).
\end{equation}
The r.h.s. of Eq. (\ref{eq33}) is the Lagrangian density in the
Einstein frame.
 It is interesting to stress that for the potential
\begin{equation}\label{eq34}
V(\sigma)=\frac{\bar{\Lambda}}{36}\sigma^4,
\end{equation}
where $\bar{\Lambda}$ is a constant, we obtain
$\bar{V}=\bar{\Lambda}$  and the action in Einstein frame is
reduced exactly to the Hilbert-Einstein action with a cosmological
constant $\bar{\Lambda}$. The corresponding potential in the
Jordan frame with Brans-Dicke action (\ref{eq27}) is
\begin{equation}\label{eq35}
V(\phi)=4\bar{\Lambda}\phi^2,
\end{equation}
which converts the action into a gravity theory non-minimally
coupled with a massive scalar field with an squared mass scale of
the order of cosmological constant.

\section{Discussions and Conclusions}
\label{5}

Summarizing, we have considered four actions: metric-Jordan
(\ref{eq15'}),
 Palatini-Jordan (\ref{eq27}),  metric-Einstein (\ref{eq27'})  and  Palatini-Einstein
(\ref{eq33}).  Jordan and Einstein frames, i.e. the actions
(\ref{eq15'}) and (\ref{eq27'}), are  related by a conformal
symmetry. In this case,  the appearance of a kinetic term is the
relevant feature. The actions (\ref{eq27}) and (\ref{eq33}) are
also related by a conformal symmetry. However,  in this case, the
kinetic term is not present. In other words, the conformal
symmetry between Jordan and Einstein frames in metric and Palatini
formalisms corresponds to the appearance or the vanishing of a
kinetic term. On the other hand, comparing (\ref{eq15'}) with
(\ref{eq27}) reveals  that the transition from metric-Jordan
action (\ref{eq15'}) to Palatini-Jordan action (\ref{eq27})
requires the appearance of a kinetic term, while the transition
from metric-Einstein action (\ref{eq27'}) to the Palatini-Einstein
action (\ref{eq33}) requires the vanishing of kinetic term. This
fact could have a deep dynamical interpretation. We have already
learned about the conformal transformations relating Jordan with
Einstein frames and Palatini with metric formalisms. Jordan and
Einstein frames are {\it dynamically} equivalent from the
conformal symmetry viewpoint. Although the  metric and Palatini
formalisms are connected through a conformal transformation
(\ref{eq19}), they apparently do not seem to be dynamically
equivalent. Metric-Jordan action differs from Palatini-Jordan
action with a dynamical {\it advanced} kinetic term. In the same
way, metric-Einstein action differs from Palatini-Einstein action
with a dynamical {\it retarded} kinetic term. However, the
Palatini-Jordan action, when reduced to the Palatini-Einstein
action, takes the same form as the metric-Jordan action, namely it
becomes of the O'Hanlon type action where  dynamics is completely
endowed by the self-interacting potential. On the other hand,
metric-Einstein action and Palatini-Jordan action represent the
same dynamical features because both have a dynamical kinetic term
plus a potential. In conclusion, for each map between Jordan and
Einstein frames, there exists a corresponding map between Palatini
and metric formalisms. In the same way, for each map connecting
two O'Hanlon type actions, namely metric-Jordan and
Palatini-Einstein action, there exists a map which connects
Palatini-Jordan action with metric-Einstein action. In conclusion,
the dynamical content of Palatini and metric formalism is exactly
the same.

Beside the mathematical consistency of  Einstein
relativity versus more general theories,  it is important to point
out the physical motivations of these approaches.  In general,
scalar fields are introduced to solve the shortcomings of the
Standard Cosmological Model (addressed by the inflationary
paradigm \cite{linde}) or issues as dark matter and dark energy
(addressed by quintessence models, induced-matter theory, etc.
\cite{copeland}). Several results point out  that a scalar field
should come from a Kaluza-Klein theory than a 4D theory, and the
Brans-Dicke theory could appear obsolete in this picture. For
example Coley et al. have proved that all results of 4D
Brans-Dicke theory can be obtained more easily from a 5D
Kaluza-Klein theory (see e.g. \cite{coley1,coley2}) while in
\cite{bellini1,bellini2} it is proved that extending General
Relativity in 5D can easily give rise to mechanisms  capable of
generating inflation and dark energy behavior.

Beside these fundamental physics motivations,  scalar
fields  represent the further degrees of freedom that
gravitational interaction can present once we do not strictly
consider General Relativity as the only possible theory of
gravity. In fact, relaxing the hypothesis that the gravitational
action is only the Hilbert-Einstein one, it is widely recognized
that $f(R)$-theories or theories constructed by other curvature
invariants could address inflation, dark energy and dark matter
problems \cite{mauro,farasot,libroSaFe,f(R)-cosmo,f(R)-cosmo1,f(R)-cosmo2,f(R)-cosmo3}. The fact that
Jordan-Einstein frames and Palatini-metric formalisms have the
"same" dynamical content means that the "scalar field" can be
represented in several ways.  However an open question remains: is
it a genuine new ingredient at fundamental level (e.g. the Higgs
Boson or a Kaluza-Klein field) or is it an average effect induced
by geometry?
 Very likely the forthcoming experimental results at LHC (CERN)
 could give hints to address this issue.

\section*{Acknowledgment}
SC is supported by INFN,  Sez. di Napoli. FD is supported by
``Research Institute for Astronomy and Astrophysics of Maragha
(RIAAM)'', Iran. FD appreciates R.I.A.A.M for  hospitality.

\end{document}